\pdfoutput=1
\documentclass[11pt]{article}

\usepackage[preprint]{acl}

\usepackage{times}
\usepackage{latexsym}

\usepackage[T1]{fontenc}
\usepackage[utf8]{inputenc}

\usepackage{microtype}

\usepackage{inconsolata}

\usepackage{graphicx}

\usepackage{hyperref}
\usepackage{url}
\usepackage{amsmath}
\usepackage{multirow}
\usepackage{hyperref}
\usepackage{subfigure}
\usepackage{booktabs}
\usepackage{threeparttable}
\usepackage{amssymb}
\usepackage{gensymb}
\usepackage{enumitem}
\usepackage{color}
\usepackage{bbding}
\usepackage{algorithm}
\usepackage{algorithmic}
\title{USM-VC: Mitigating Timbre Leakage with Universal Semantic Mapping Residual Block for Voice Conversion}

\author{Na Li\thanks{Equal contribution.} \\
  Tencent AI LAB \\
  {\small\texttt{nenali@tencent.com}} \\ \And
  Chuke Wang\footnotemark[1] \\
  Peking University \\
  {\small\texttt{chuke@stu.pku.edu.cn}} \\ \And
  Yu Gu \\
  Tencent AI LAB \\
  {\small\texttt{colinygu@tencent.com}} \\ \And
  Zhifeng Li \\
  {\small\texttt{zhifeng0.li@gmail.com}} \\  
}

\begin{document}

\maketitle
%\footnote[0]{*Equal contribution.}
\begin{abstract}
Voice conversion (VC) transforms source speech into a target voice by preserving the content. However, timbre information from the source speaker is inherently embedded in the content representations, causing significant timbre leakage and reducing similarity to the target speaker. To address this, we introduce a \textbf{Universal Semantic Matching (USM)} residual block to a content extractor. The residual block consists of two weighted branches: 1) universal semantic dictionary based Content Feature Re-expression (CFR) module, supplying timbre-free content representation. 2) skip connection to the original content layer, providing complementary fine-grained information. In the CFR module, each dictionary entry in the universal semantic dictionary represents a phoneme class, computed statistically using speech from multiple speakers, creating a stable, speaker-independent semantic set. We introduce a CFR method to obtain timbre-free content representations by expressing each content frame as a weighted linear combination of dictionary entries using corresponding phoneme posteriors as weights. Extensive experiments across various VC frameworks demonstrate that our approach effectively mitigates timbre leakage and significantly improves similarity to the target speaker. The code and pretrained models are available at \url{https://github.com/DisplayVoiceDemo/USM-voice-conversion}
\footnote{The demo is available at \url{https://displayvoicedemo.github.io/USM/}}.
% Timbre-free semantic representation is also crucial for other speech generation tasks such as text-to-speech (TTS), song generation, and singing voice conversion.
\end{abstract}

\section{Introduction}
\label{intro}
Content representations play a role of determining the linguistic content of the generated audio in speech generation tasks such as text-to-speech (TTS), song generation, and singing voice conversion. However, content representations often also contain timbre, prosody, and other information, which can pose significant challenges for tasks that aim to generate audio with specific timbre characteristics. A typical example is voice conversion which directly uses content representation as condition to generate speech in a target speaker's voice, the timbre information inherited in source content representations dramatically decrease the similarity between the generated speech and the target speaker. This paper focuses on the Voice Conversion (VC) task and investigates how to develop timbre-independent content representations.

Most research efforts in the field of voice conversion focus on disentangling timbre from content representations. These approaches aim to achieve information disentanglement by employing complex feature engineering \citep{li2023freevc, qian2019autovc, streamvc23, choi2024dddm}, specialized network architectures and training strategies \citep{wang2021vqmivc, wang2023lm, ju2024naturalspeech, lajszczak2024base}, normalization techniques \citep{chou2019one}, or data augmentation strategies \citep{li2023freevc, anastassiou2024seed}. However, these methods still suffer from timbre leakage and struggle to maintain timbre similarity with the target speaker. 
%such approaches are not universally applicable or challenging to reproduce. 

In this paper, we address the issue of timbre leakage from a novel perspective. The fundamental cause of timbre leakage is that the timbre information of the source speaker is inherently embedded in the representations of the source speech. This leads us to the following question: What type of content representation can exclude timbre information? Previous studies \citep{polyak2021speech,van2020vector,huang2021any} have demonstrated that discrete speech units, derived from clustering self-supervised representations, can function as timbre-free content units. This is because the discretization process introduces an information bottleneck that effectively separates content from timbre information. Inspired by this, we introduce a Universal Semantic Matching (USM) residual block to the content extractor. The residual block consists of a universal semantic dictionary based Content Feature Re-expression (CFR) module and a weighted skip connection to the content layer. The USM block needs an offline construction of a universal semantic dictionary composed of discrete entries. Each entry in the Universal Semantic Dictionary is calculated as a weighted combination of content representations from multiple speakers, distinguishing it from discrete speech units obtained through clustering. The CFR module re-expresses each content feature from the source speaker as a weighted combination of the entries in the Universal Semantic Dictionary. The weighted skip connection to the content layer provides complementary contextual information for the timbre-free content representations extracted from the CFR module. 

We apply the new content representations derived from the USM residual block to various voice conversion (VC) frameworks, including language model-based zero-shot VC, diffusion model-based one-shot VC, and Variational Inference with adversarial learning for end-to-end Text-to-Speech (VITS)-based any-to-many VC. This application results in substantial improvements in similarity and speech naturalness compared to the original representation. 
\begin{figure*}[ht]
    \centering
    \includegraphics[scale=0.67]{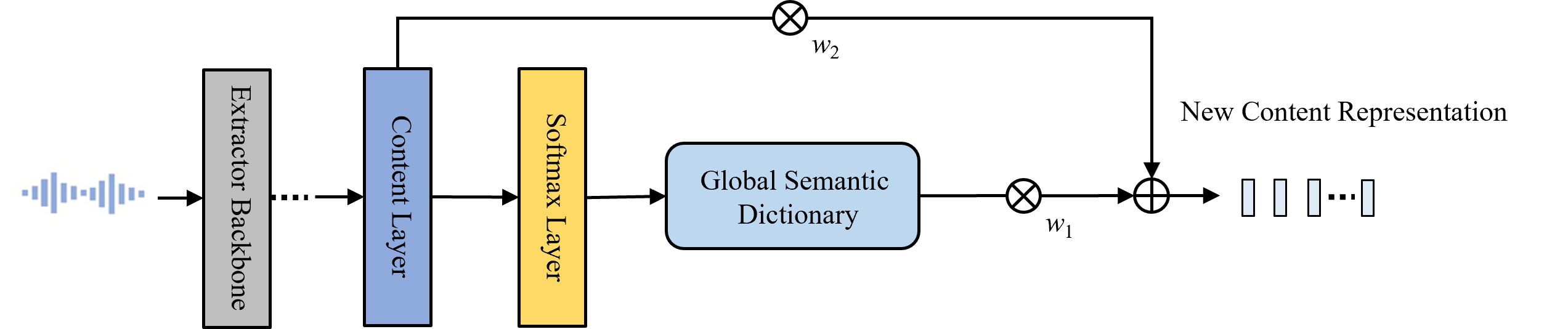}    
    \caption{Illustration of the proposed USM residual block.}    
    \label{usm}
\end{figure*}
In conclusion, our contributions are as follows: 

1. We propose a novel Universal Semantic Matching (USM) residual block to extract new content representation for voice conversion. In USM block which consists of a universal semantic dictionary based Content Feature Re-expression (CFR) module and a weighted skip connection to the content layer, an offline universal semantic dictionary is first constructed by utilizing content representations from various speakers. Each entry in the dictionary provides a stable, timbre-independent representation of a specific phoneme class or speech unit. Based on this dictionary, the Content Feature Re-expression (CFR) module aims to express each frame of the original content features as a linear weighted combination of the dictionary entries, yielding novel timbre-independent representations that are highly beneficial for voice conversion. The weighted skip connection to the content layer further provides complementary contextual information for the timbre-free content representations.

2. Compared to widely used information decoupling methods that rely on complex network architectures and intricate training strategies to mitigate timbre leakage, our approach is inherently free of timbre information. Furthermore, it is easier to implement and significantly reduces computational complexity, time complexity, and model size.

3. We conduct extensive experiments across various VC frameworks. The results show that our method not only outperforms existing state-of-the-art approaches but also demonstrates strong generalization capabilities, making it potentially applicable to other speech generation tasks.

4. This work establishes a new paradigm for tackling the complex timbre leakage problem, achieving highly expressive results across various settings while significantly reducing computational complexity, time complexity, and model size. We believe it offers valuable insights for future research and makes a substantial contribution to the ongoing development of this field.

\section{Related Works}
Content representations are typically extracted from the bottleneck layer of supervised pre-trained phoneme-posteriorgram (PPG) models \citep{streamvc23, liu2021any, kovela2023any} or an intermediate layer of self-supervised models like HuBERT \citep{hsu2021hubert} and WavLM \citep{chen2022wavlm}. However, since the training audio inherently contains content and timbre information, these representations inevitably include undesirable timbre, reducing the similarity between the converted speech and the target speaker. 

To address the above issue, discrete speech units \citep{van2020vector,huang2021any} are introduced into voice conversion. However, discrete speech units may lack some linguistic content, and distance-based discretization may cause ambiguous or noisy representations to be assigned to incorrect nearby units, resulting in mispronunciation. \citep{van2022comparison} proposes to replace discrete speech units with soft speech units. Though intelligibility and naturalness improvements are achieved, such representations lose the discriminability between adjacent frames and rich contextual information, the converted speech exhibits issues such as unclear pronunciation or unnatural prosody. 

Most recent voice conversion (VC) methods focus on decoupling timbre and content information. These methods can be divided into several categories: 1) information bottlenecks. FreeVC \citep{li2023freevc} disentangles content information by imposing an information bottleneck on WavLM features to reduce the timbre information contained in the content representation. In VQMIVC \citep{wang2021vqmivc}, vector quantization (VQ) is employed as a discretization strategy to impose an information bottleneck for content encoding. 2) specialized network designs and training strategies. Decoupled denoising diffusion models (DDDMs) \citep{choi2024dddm} utilize multiple attribute denoisers to address the challenges of disentangling and controlling speech attributes for VC tasks. Mutual information (MI) minimization is introduced in \citep{wang2021vqmivc} to achieve information disentanglement. 3) data augmentation. Spectrogram-resize-based data augmentation is proposed in \citep{li2023freevc} to enhance content representation. Although these decoupling methods have achieved some success in separating timbre and content information, timbre is inherently embedded in speech, making decoupling challenging. Consequently, timbre leakage is often unavoidable. Additionally, decoupling networks tend to be quite complex, significantly increasing the burden on the system. 

Unlike decoupling methods, KNN-VC \citep{baas2023voice} uses the K-Nearest Neighbors algorithm to replace each frame of the source speech's content representation with the nearest neighbor from the target speaker's representations for voice conversion. However, it struggles with identifying accurate neighbors for noisy content, causing pronunciation issues, and its zero-shot performance is limited by the amount of available target speech.

\begin{figure}[htb]
  % \centering
  % \vspace{-0.5em}
    \begin{minipage}[b]{.48\linewidth}
      \centering
      \centerline{\includegraphics[width=3.04cm,scale=0.5]{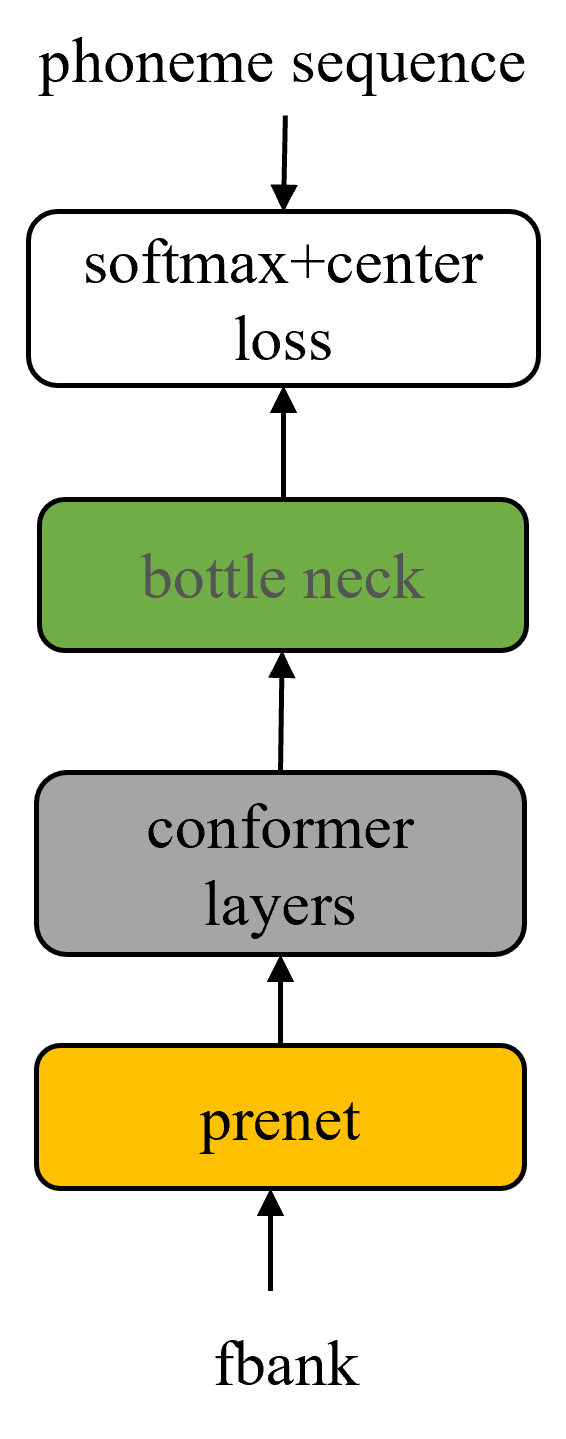}}
    %  \vspace{1.5cm}
      \centerline{(a) PPG Extractor}\medskip
    \end{minipage}
    \begin{minipage}[b]{.48\linewidth}
      \centering
      \centerline{\includegraphics[width=3.25cm,scale=0.5]{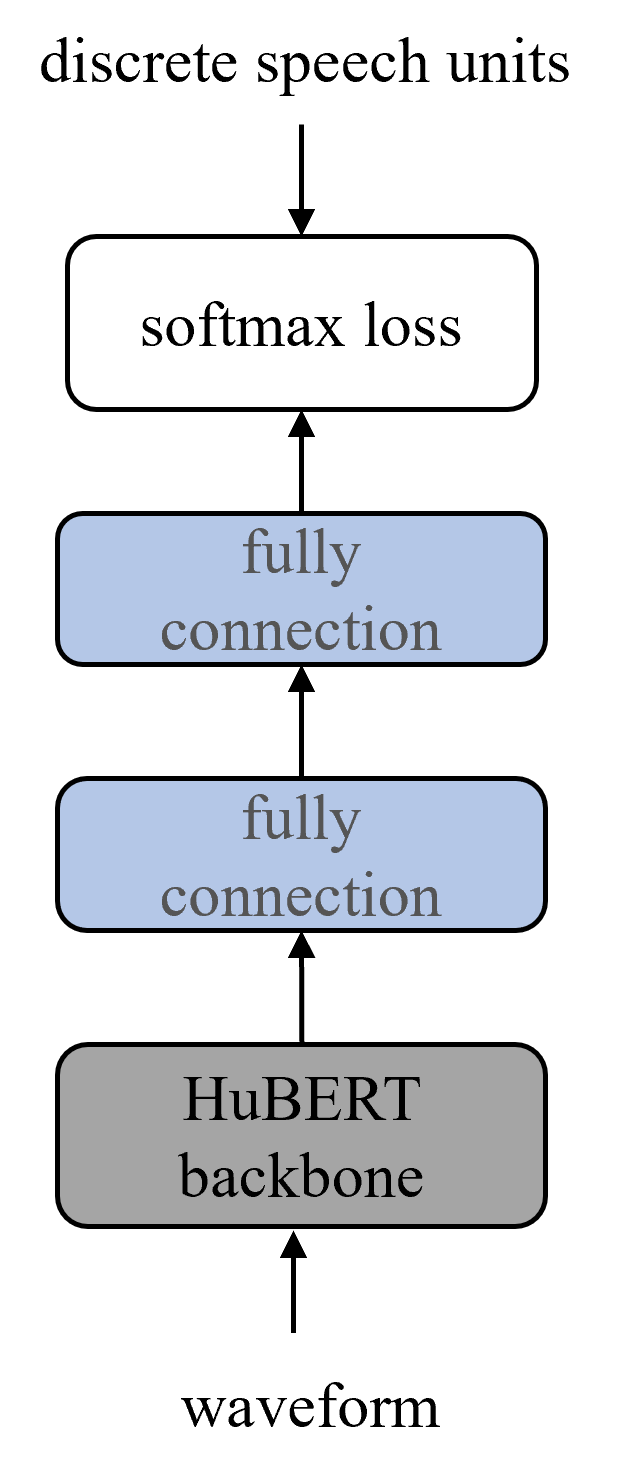}}
    %  \vspace{1.5cm}
      \centerline{(b) HuBERT Extractor}\medskip
    \end{minipage}     
  % \includegraphics[scale=0.6]{ppg.png}
  % \vspace{-0.5em}
  \caption{Illustration of supervised PPG extractor and self-supevised HuBERT extractor.}\label{extractor}
\end{figure}

\section{Method}
\label{method}
In this Section, we introduce the proposed Universal Semantic Matching (USM) residual block. Fig.\ref{usm} shows the process of obtaining the new content representation with USM. In USM, we first construct a universal semantic dictionary that can be applied to both self-supervised and supervised representations. Each entry in the semantic dictionary is computed as a weighted combination of content representations from different speakers. The posterior distribution of phoneme units, extracted from the softmax layer of a content extractor, is used as the combination weights. Based on the universal semantic dictionary, the Content Feature Re-expression (CFR) module represents each frame of the original content features as a linear weighted combination of the dictionary entries. The weighted skip connection further provides complementary contextual information for the timbre-free content representations extracted from the CFR module.

\subsection{Content Extractor}
\label{sec:ppg model}
Supervised PPG or self-supervised model such as HuBERT can be used to extract content representations or soft speech units, as shown in Fig.\ref{extractor}.
The PPG model is trained with phonetic alignments with acoustic features extracted from the HMM-DNN \citep{povey2018semi} model using Kaldi Tookit\footnote{https://github.com/kaldi-asr/kaldi}. The architecture of PPG model is shown as in Fig.\ref{extractor}(a). We employ cross entropy objective to train the model, and introduce the center loss \citep{wen2016discriminative} as an auxiliary optimization to improve the robustness of the content representation extracted from the bottle neck layer. 

For the HuBERT model, we frozen the backbone and used the seventh transformer layer to extract content features. To learn discrete speech units, we apply k-means clustering to the content features extracted from speech of different speakers. For the learning of soft speech units, two fully connection layers are added after the seventh transformer layer. We fine-tune the whole model shown in Fig.\ref{extractor}(b) to predict the corresponding discrete speech units.

\subsection{Construction of Universal Semantic Dictionary}
\label{sec:dict construct}
The universal semantic dictionary is computed using a development audio set with $S$ speakers. Firstly, we define the calculation methods for zero-order and first-order statistics, respectively. Specifically, the zero-order statistics are calculated using the weights of the classification layer of a content extractor. The first-order statistics are computed by a linear weighted combination of the content representations. Denote $\mathbf {x}_{i,j,t} \in \mathbb{R}^{d\times1}$, where $i=1,..S, j=1,...N_{i}, t=1,..T_{ij}$, is the $t$-th frame content representation from $i$-th speaker's $j$-th utterance. Let $\gamma_{i,j,t}^{k}$ denote the posterior probability that the current frame belongs to the $k$-th phoneme class or discrete speech unit. The speaker-independent zero-order statistics are computed according to Eq.\ref{Eq:1}. Then the speaker-independent first-order statistics $\mathbf {m}_{k} \in \mathbb{R}^{d\times1}$ are obtained using Eq.\ref{Eq:2}. Finally, the universal semantic dictionary $\mathbf M_{g} \in \mathbb{R}^{d\times K}$ of size $K$ can be represented as Eq.\ref{Eq:3}, where $K$ is the total number of phoneme classes or discrete speech units.

\begin{equation}
    n_{k}=\sum_{i,j,t}\gamma_{i,j,t}^{k}
\label{Eq:1}
\end{equation}

\begin{equation}
    \mathbf {m}_{k}=\dfrac{1}{n_{k}}\sum_{i,j,t}\gamma_{i,j,t}^{k}\mathbf {x}_{i,j,t}
\label{Eq:2}
\end{equation}

\begin{equation}
    \mathbf {M}_{g}=[\mathbf {m}_{1},\mathbf {m}_{2},...\mathbf {m}_{K}]
\label{Eq:3}
\end{equation}

\subsection{Content Feature Re-expression (CFR) Using Universal Semantic Dictionary}
\label{sec:extraction}
Given the $t$-th frame original content representation $\mathbf {x}_{i,j,t}$ and the corresponding posterior probability $\mathbf {p}_{i,j,t} \in \mathbb{R}^{K\times1}$ associated to each phoneme class or discrete speech unit. A new timbre-independent content representation $\bar {\mathbf {x}}_{i,j,t}$ can be obtained according to Eq.\ref{Eq:7}. 
% For any-to-many VC task, a target speaker-related content representation $\hat {\mathbf {x}}_{i,j,t}$ can be computed using Eq.\ref{Eq:8}.
\begin{equation}
    \bar {\mathbf {x}}_{i,j,t}= \mathbf {M}_{g}\mathbf {p}_{i,j,t}
\label{Eq:7}
\end{equation}

\subsection{Content Representation From USM Residual Block}
The USM representation $ \hat {\mathbf {x}}_{i,j,t}$ extracted from USM residual block is a weighted linear combination of the original content representation $\mathbf {x}_{i,j,t}$ and the timbre-independent content representation $\bar {\mathbf {x}}_{i,j,t}$, as shown in Eq.\ref{Eq:8}
\begin{equation}
    \hat {\mathbf {x}}_{i,j,t}= w_{1}\bar {\mathbf {x}}_{i,j,t}+w_{2}{\mathbf {x}}_{i,j,t}
\label{Eq:8}
\end{equation}
where $w_{1}+w_{2}=1$. $w_{1}$ controls the contribution of the timbre-free content representation, $w_{2}$ regulates the contribution degree of fine-grained contextual information within the original representation.

\section{Applying USM Across Different Voice Conversion Frameworks}
\subsection{VITS based Any-to-Many Voice Conversion}
\label{sec:vits vc}
Previous VC systems utilize a two-stage reconstruction pipeline \citep{qian2019autovc,liu2021any}. Initially, a conversion model transforms source acoustic features into the target speaker's domain, followed by a vocoder converting these features into waveform in the second stage. VITS, a one-stage model capable of both TTS and VC, connects the two stages via latent variables of a conditional variational autoencoder (CVAE), thereby reducing feature mismatch. Furthermore, adversarial training improves the quality of the reconstructed waveform. 
% This paper tries to use CFR to improve the speaker similarity and maintain the high quality of the generated waveform.

Given an input utterance, speech unit/phoneme posterior probability $\mathbf {p}_{i,j,t}$ is extracted from a content extractor. Then timbre-independent content representations are calculated according to Eq.\ref{Eq:7}. The weighted skip connection further introduces complementary contextual information for the timbre-free content representations. A look-up table (LUT) is employed as the speaker identity indicator. Given the content feature and target speaker indicator, VITS model can generate speech with target timbre. For any-to-many voice conversion task, it is feasible to construct a speaker-dependent semantic dictionaries for individual target speaker within the training set, the resulting new content representation via CFR will contain target timbre information which is beneficial for the conversion task.

\subsection{Language Model based Zero-Shot Voice Conversion}
\label{sec:lm vc}
Large language models (LLMs) have demonstrated great progress in natural language generation. With a sufficient model size LLMs emerge powerful in-context learning abilities that can handle unseen tasks with a prompt in a zero-shot or few-shot manner \citep{wei2022emergent}. Moreover, the simple yet effective next-token prediction task of LLMs makes it easy to apply LLMs on voice conversion \citep{wang2023lm}, as long as the data can be converted to discrete speech tokens. An intuitive approach is to follow AudioLM \citep{borsos2023audiolm} where speech was tokenized into semantic and acoustic tokens by HuBERT and a neural codec, respectively. Subsequently, generate the target acoustic tokens by autoregressive prediction of the next token, conditioned on the semantic tokens and the acoustic tokens of the prompt audio. While, semantic tokens lose much rich linguistic information, resulting in hard contextual learning and poor audio quality.

In this paper, we also investigate the effects of the proposed USM block on zero-shot voice conversion using a decoder-only language model with a neural codec. Similar to VALL-E \citep{wang2023neural}, the model predicts target codec tokens hierarchically based on a sequence of codec tokens from the prompt speech segment and content features extracted from the source speech through USM block. In this process, the codec tokens of the first layer of RVQ are predicted by an AR transformer, while the tokens of the remaining layers are predicted by a NAR transformer.

\subsection{Diffusion Model based One-Shot Voice Conversion}
\label{sec:diffusion vc}
The diffusion models have achieved remarkable performance on VC tasks \citep{liu2021diffsvc,lu2024comosvc,chen2024ldm}, producing natural speech with high similarity to the target timbre. To verify the universality of the USM block, we also verified the effectiveness of the USM block on the diffusion model. Diffusion model is adopted as the probabilistic model which fits the distribution of mel-spectrogram. We use the EDM method \citep{karras2022elucidating} to train the diffusion model. In the training stage, the content representations and the corresponding speaker embedding are encoded into hidden embeddings. These embeddings are concatenated and served as the conditional input
for the network. In the inference process, conditioned on the source speaker's content representation and target speaker embedding, we iteratively
sample the target mel-spectrogram from a Gaussian noise. The generated mel-spectrogram can be further rendered to audio by using a pre-trained vocoder.

\section{Experiment}
\label{sec:exp}
In order to verify the effectiveness of the USM block, we compared the effects of different content representations on different VC tasks, including the original content representations, softmax speech units, and the proposed USM representation obtained from different extractors.

\subsection{Evaluation Metrics}
We assess the quality of the converted audios utilizing three objective metrics: the F0 Pearson Correlation (FPC), the Speaker Similarity (SSIM), and word error rate (WER). For FPC, we calculated the L1 distance between the log-scale ground-truth and the predicted F0 in the HAG. To obtain the ground-truth F0, we compute the mean of F0 values of the target speaker and source speech, denoted as $\bar{f}_{0}^{tar}$ and $\bar{f}_{0}^{src}$ respectively. The ground truth F0 is obtained according to $f_{0}^{src} \times \bar{f}_{0}^{tar} / \bar{f}_{0}^{src}$. SSIM is computed through cosine similarity using speaker embeddings derived from an Automatic Speaker Verification model. WER is calculated using a pre-trained automatic speech recognition (ASR) model. For subjective evaluations, we conduct a 5-point Mean Opinion Score (MOS) test, ranging from 1 (bad) to 5 (excellent). A total of 10 volunteers are recruited for the listening test, where they provide ratings for both the Naturalness Mean Opinion Score (NMOS) and the Similarity Mean Opinion Score (SMOS). A confidence interval 0f $95\%$ is reported for MOS. For simplicity, in some experiments, we adopt UTMOS \citep{saeki2022utmos} instead of NMOS as an objective metric for naturalness.

\begin{table*}[ht]
{
\small
\caption{\label{tbl}Performance of different content representations on subjective metrics (NMOS, SMOS) and objective metrics (SSIM, FPC, WER) for VITS based any-to-many VC task. $\text{USM}^{*}$ denotes the representation that incorporates speaker-dependent content representation by CFR using speaker-dependent semantic dictionary for each target speaker.}
\begin{center}
\begin{tabular}{cc|cccccc} \hline
 \multicolumn{2}{c|}{Content Representation} & NMOS$\uparrow$  & SMOS$\uparrow$ & SSIM$\uparrow$ & FPC$\uparrow$ & WER$\downarrow$ \\ \hline
  \multirow{4}{*}{PPG} & BNF &  $4.012\pm0.092$ & $3.051\pm0.091$ & 0.601 & 0.585 & 2.285  \\ 
 & S-Unit &  $3.791\pm0.093$ & $3.523\pm0.107$ & 0.765 & 0.601 & 4.596 \\ 
 & \textbf{USM} & $\textbf{4.153}\pm0.096$ & $3.832\pm0.093$ & 0.748 & 0.781 & \textbf{2.102} \\
 & $\textbf{USM}^{*}$  & $4.013\pm0.101$ & $\textbf{4.112}\pm0.102$ & \textbf{0.796} & \textbf{0.785} &  2.262\\ \hline
 \multirow{4}{*}{HuBERT} & MLF  & $3.959\pm0.094$ & $2.879\pm0.102$ & 0.403 & 0.561  & 2.345 \\ 
 & S-Unit & $3.653\pm0.101$ & $3.654\pm0.095$ & 0.773 & 0.639 & 4.895 \\
 & \textbf{USM} &  $3.932\pm0.096$ & $3.701\pm0.093$ & 0.732 & 0.767 & \textbf{2.193} \\ 
  & $\textbf{USM}^{*}$  & $\textbf{4.005}\pm0.091$ & $\textbf{4.166}\pm0.101$ & \textbf{0.808} & \textbf{0.793} & 2.234 \\ \hline
\end{tabular}
\end{center}}
%\vspace{-0.5cm}
\end{table*}

\begin{table*}[htbp]
{
\small
\caption{\label{tb3}Performance of different content representations on subjective metrics (NMOS, SMOS) and objective metrics (SSIM, FPC, WER) for language model based zero-shot VC task.}
% \vspace{-s0.5cm}
\begin{center}
\begin{tabular}{cc|ccccc} \hline
 \multicolumn{2}{c|}{Content Representation}  & NMOS$\uparrow$  & SMOS$\uparrow$ & SSIM$\uparrow$ & FPC$\uparrow$ & WER$\downarrow$ \\ \hline
  \multirow{3}{*}{PPG} & BNF  &  $4.215\pm0.091$ & $3.268\pm0.081$ & 0.641 & 0.653 & 2.153  \\ 
 & S-Unit & $3.618\pm0.091$ & $3.421\pm0.089$ & 0.711 & 0.632 & 4.446 \\ 
 & \textbf{USM} &  $\textbf{4.246}\pm0.088$ & $\textbf{3.845}\pm0.094$ & \textbf{0.751} & \textbf{0.765} & \textbf{2.133} \\ \hline 
\multirow{3}{*}{HuBERT} & MLF  & $4.306\pm0.087$ & $3.254\pm0.088$ & 0.624 & 0.612  & \textbf{1.991} \\ 
 & S-Unit  & $3.421\pm0.096$ & $3.312\pm0.097$ & 0.683 & 0.603 & 5.526 \\ 
 & \textbf{USM} & $\textbf{4.314}\pm0.093$ & $\textbf{3.823}\pm0.097$ & \textbf{0.741} & \textbf{0.756} &  2.115 \\ \hline
\end{tabular}
\end{center}}
%\vspace{-0.5cm}
\end{table*}

\subsection{Datasets}
\label{sec:dataset}
WenetSpeech \citep{zhang2022wenetspeech} and Gigaspeech \citep{chen2021gigaspeech} are used to train PPG model introduced in Sec.\ref{sec:ppg model}. We choose the open source vocabulary, BigCiDian\footnote{https://github.com/speechio/BigCiDian}, as our lexicon.

Our experiments were carried out on VCTK \citep{vctk} and LibriTTS \citep{zen2019libritts}. Only VCTK is used for training VITS based systems. All recordings are resampled to 24 kHz. The whole train set of LibriTTS is used to train the diffusion and language models. For subjective evaluation tests, the test audio samples are selected from the test set of LibriTTS corpus. We randomly choose 20 target speakers and 10 test audio samples for each target speaker for VITS and diffusion model based frameworks. For language model based zero-shot VC, we randomly select 100 prompt audio clips with less than 10s duration from 50 unseen speakers in the VCTK corpus and 5 test audio samples from the rest speakers for each prompt. For calculating objective metrics, 100 test audios are randomly selected for each target speaker and 20 test audios are selected for each prompt.

\subsection{Experimental Setup}
\label{sec:setup}
\noindent\textit{\textbf{PPG Extractor:}} In the training stage, the input audio was augmented with noise, music, and reverberation. The input spectral features are 80-dimensional log mel-spectrograms with 10ms hop size and 25 ms window size. The stem of the model is built from a pre-net (two linear layers with dropout), followed by a stack of seven conformer blocks with 4 attention heads, a kernel size of 15, a hidden size of 256, and a filter size of 2048. The output size of the bottle neck layer is 256. The semantic dictionary with $600$ entries of 256-dimensional is calculated using 100,000 randomly selected audio samples from 2,311 speakers in the train set of LibriTTS.

\begin{table*}[htbp]
{
\small
\caption{\label{tb4}Performance of different content representations extracted from PPG model on subjective metrics (NMOS, SMOS) and objective metrics (SSIM, FPC) for diffusion model based VC task.}
% \vspace{-s0.5cm}
\begin{center}
\begin{tabular}{c|ccccc} \hline
 \multicolumn{1}{c|}{Content Representation}  & NMOS$\uparrow$  & SMOS$\uparrow$ & SSIM$\uparrow$ & FPC$\uparrow$ & WER$\downarrow$\\ \hline
BNF  & $4.002\pm0.088$ & $3.241\pm0.086$ & 0.652 & 0.627 & \textbf{1.338} \\
S-Unit & $3.986\pm0.097$ & $3.835\pm0.092$ & \textbf{0.766} & 0.632 & 3.084  \\ 
\textbf{USM} & $\textbf{4.146}\pm0.086$  & $\textbf{3.962}\pm0.093$ & 0.759 & \textbf{0.701} & 1.575 \\ \hline 
\end{tabular}
\end{center}}
%\vspace{-0.5cm}
\end{table*}

\noindent\textit{\textbf{HuBERT Extractor:}} The output of the seventh transformer layer in HuBERT-Base \citep{hsu2021hubert} is used as the original self-supervised content representation. For discrete speech units, we apply k-means clustering to content representations. we adopt $K$ clusters and estimate their means on a set of 200,000 audio samples randomly selected from 2,311 speakers from the train set of LibriTTS. To obtain soft speech units, we froze the HuBERT backbone and add two linear projection layers with 256 hidden states after the seventh layer, followed by a classification layer, whose target is the label of the corresponding k-means clustering center. The semantic dictionary with $K$ entries of 768-dimensional is constructed using the same set used for estimating the PPG extractor-related semantic dictionary. 

\noindent\textit{\textbf{VITS System:}} The model architecture is similar to the open-source RVC project\footnote{https://github.com/RVC-Project/Retrieval-based-Voice-Conversion-WebUI}. Specifically, the posterior encoder utilizes non-causal WaveNet residual blocks \cite{prenger2019waveglow}. The prior encoder consists of a 6-layer transformer with 2 attention heads. Normalizing flows, which conditions on speaker embedding is adopted to improve the complexity of prior distribution. The decoder follows the original configuration of the HIFI-GAN \citep{kong2020hifi} in VITS. We adjust the model's hyperparameters to suit the generation of 24kHz audio, resulting in 36M parameters in total. More experimental details can be found in Appendix \ref{apx:1}.

\noindent\textit{\textbf{Language Model based VC System:}} The architecture of the codec language model is similar to VALL-E \citep{wang2023neural} where both the AR and NAR models are a 12-layer transformer with 12 attention heads and 768-dimensional token embeddings, sinusoidal positional embeddings, 3072-dimensional feed-forward layers, and a dropout rate of 0.1. The model size is 227M. The encoder and decoder of the pre-trained codec model follows Hifi-codec \citep{yang2023hifi}, except that the number of quantization layer of RVQ is set to 4 with the group number set to 1. More experimental details can be found in Appendix \ref{apx:2}.

\noindent\textit{\textbf{Diffusion Model based VC System:}} The training pipeline is a variant version of the CoMoSVC project\footnote{https://github.com/Grace9994/CoMoSVC}. We only trained the teacher model for our experiments. The 256-dimensional speaker embedding extracted from a pre-trained speaker verification model was used as the condition to control the generated timbre. Unlike the CoMoSVC project, We did not incorporate pitch and loudness as conditional inputs into the network. For the model architecture, we replaced the original WaveNet with the flow matching decoder from the StableTTS project\footnote{https://github.com/KdaiP/StableTTS}. The decoder consists of 12 Convolution Transformer blocks modified from Hierspeech++ \citep{lee2023hierspeech++}. Each Convolution Transformer block contains a FiLM layer \citep{perez2018film}, three ConvNeXt blocks with a hidden size of 768, a filter size of 2048 and a kernel size of 7, and a DiT block with 8 attention heads, a kernel size of 3, a hidden size of 768, and a filter size of 768. The output of the decoder is the predicted 80-dimensional log mel-spectrogram. The total size of the model is 287M. More experimental details can be found in Appendix \ref{apx:3}.

\subsection{Results and Analysis}
\label{sec:ra}
In the following experiments, MLF denotes the features extracted from the 7-th layer of the HuBERT model. BNF is the bottle neck feature extracted from the PPG model. S-Unit is the soft speech units. The number of clusters $K$ for k-means clustering is set to 4096 for USM.
\subsubsection{Results of USM for Different VC Frameworks}
\label{sec:dif-vcs}
\noindent\textit{\textbf{Results of VITS based VC Systems:}}
Comparison of different content representations in any-to-many voice conversion based on VITS architecture is shown in Table \ref{tbl}. For USM, the values of $w_{1}$ and $w_{2}$ are 0.8 and 0.2, respectively. $\text{USM}^{*}$ incorporates speaker-dependent content representation with weight $w_{3}$. For $\text{USM}^{*}$, the values of $w_{1}$, $w_{2}$ and $w_{3}$ are 0.2, 0.6 and 0.2, respectively. The impact of different weight combinations on the conversion effect is shown in Appendix \ref{apx:4}.

Regarding the quality of the generated speech, we can observe that the original content representation BNF ang MLF can achieve higher NMOS and lower WER commpared to S-Unit. The reason is that the content representations directly extracted from the extractors contain more rich contextual information. However, S-Units can be considered as an approximation of discrete speech units, losing much contextual information. The USM and $\text{USM}^{*}$ demonstrate significant improvements in NMOS and WER compared to S-Unit and comparable NMOS and WER to the original content representation.

In terms of the similarity metrics (SMOS, SSIM, FPC) between the generated speech and the target speaker, the USM outperforms the original content representations. This shows that the USM effectively discarded the timbre information of the source speaker from the content representation. S-Unit has much lower SMOS compared to USM, the reason is that S-Unit has poor generation quality, which affects the subjective perception of similarity. The $\text{USM}^{*}$ achieves the highest value in all similarity metrics, demonstrating the effectiveness in incorporating speaker-dependent information in any-to-many VC task.

\begin{table}[htbp]
{
\small
\caption{\label{tb-cmp}Performance comparison between our optimal system (bold black font) for each framework and other methods. The USM representation based on the PPG model is employed.}
% \vspace{-s0.5cm}
\begin{center}
\begin{tabular}{c|ccc} \hline
 \multicolumn{1}{c|}{System}  & UTMOS$\uparrow$  & SSIM$\uparrow$ & WER$\downarrow$ \\ \hline
VQMIVC   & \multirow{2}{*}{2.372} & \multirow{2}{*}{0.358} & \multirow{2}{*}{58.332}  \\
\citep{wang2021vqmivc} & & & \\
YourTTS & \multirow{2}{*}{3.112} & \multirow{2}{*}{0.517} & \multirow{2}{*}{8.354} \\
\citep{casanova2022yourtts} & & & \\
KNN-VC  & \multirow{2}{*}{3.633} & \multirow{2}{*}{0.721} & \multirow{2}{*}{5.217} \\
\citep{baas2023voice} & & & \\
FreeVC   & \multirow{2}{*}{3.973} & \multirow{2}{*}{0.617} & \multirow{2}{*}{2.613} \\ 
\citep{li2023freevc} & & & \\
DDDM-VC  & \multirow{2}{*}{3.284} &  \multirow{2}{*}{0.632} &  \multirow{2}{*}{5.551} \\
\citep{choi2024dddm} & & & \\
\multirow{2}{*}{LM-VC}  & \multirow{2}{*}{3.982} & \multirow{2}{*}{0.641} & \multirow{2}{*}{2.153} \\
 & & & \\
\multirow{2}{*}{\textbf{LM-VC-USM}}  & \multirow{2}{*}{4.011} & \multirow{2}{*}{0.751} & \multirow{2}{*}{2.133} \\ 
& & & \\
\multirow{2}{*}{$\textbf{VITS-USM}^{*}$}  & \multirow{2}{*}{3.902} & \multirow{2}{*}{0.796} & \multirow{2}{*}{2.262} \\
& & & \\
\multirow{2}{*}{\textbf{Diffusion-S-USM}}  & \multirow{2}{*}{3.701} & \multirow{2}{*}{0.756} & \multirow{2}{*}{2.764} \\
& & & \\
\multirow{2}{*}{\textbf{Diffusion-USM}}  & \multirow{2}{*}{3.791} & \multirow{2}{*}{0.759} & \multirow{2}{*}{1.575} \\
& & & \\ \hline
\end{tabular}
\end{center}}
\vspace{-0.5cm}
\end{table}

\noindent\textit{\textbf{Results of Language Model based VC Systems:}}
The comparison of different content representations is shown in Table \ref{tb3}. For USM, the value of $w_{2}$ is set to 0.05. We observe similar phenomena across different extractors. In terms of the generated speech quality metrics NMOS and WER, the original content representations BNF and MLF exhibit better performance compared to S-Unit. Regarding similarity metrics, $\text{USM}^{*}$ shows the best performance. S-Unit is inferior to USM in both similarity and naturalness, as S-Unit contains less contextual information and requires a larger network and more data to learn effective information from the representations for better prediction of the codec token sequence. 
% When we fuse the original content representation into the USM using a very low weight, the USM-M method achieves a balanced effect in terms of naturalness and similarity. These observations are consistent with the performance on VITS-based VC systems. 

Comparing different extractors, we observe that MLF achieves a higher NMOS and a lower WER compared to BNF. In terms of similarity metrics, MLF has lower SMOS and SSIM values compared to BNF. This indicates that the intermediate layer representations from the network, such as MLF, contain richer contextual information and more timbre information compared to the upper layer representations. 

\noindent\textit{\textbf{Results of Diffusion Model based VC Systems:}}
In this section, we only compare the performance of different content representations extracted from the PPG model, due to the fact that different speech extractors exhibited similar performance patterns in the above. The value of $w_{2}$ in USM is set to 0.05. The results are shown in Table \ref{tb4}. We can see that the original representation has the lowest similarity compared to other types of content representation. The S-Unit representation achieves good naturalness, which differs from the conclusions of the above experiments. The reason may be that the ConvNeXt blocks in the model architecture can effectively enhance sound quality and have better robustness for S-Unit representations. Similarly to the above experiments, the USM achieves better similarity and comparable naturalness compared to the BNF.

\subsubsection{Comparison with Other Systems}
\label{sec:cmp}
The comparison between our optimal system for each framework and other popular decoupling methods is shown in Table \ref{tb-cmp}. LM-VC is the LM based VC system trained using BNF. Diffusion-S-USM denotes a small diffusion model with 58M parameters, which is comparable to that of DDDM-VC. Among the comparative systems, only LM-VC was implemented by ourselves, while for the other systems, publicly available pre-trained models were utilized for testing. Considering the system based on the VITS architecture, a comparison between $\text{VITS-USM}^{*}$ and FreeVC, which employs decoupling strategies, reveals that the former can achieve higher similarity while maintaining comparable naturalness to the latter. Comparing the diffusion model based systems, our Diffusion-S-USM and Diffusion-USM  significantly surpasses DDDM-VC which employs disentangled representations across all three metrics. Compared with VQMIVC, YourTTS, and KNN-VC, our USM based systems demonstrate superior performance. 

\section{Conclusions}
\label{sec:conc}
This paper proposes a novel USM residual block to mitigate timbre leakage in voice conversion. The effectiveness of the proposed method is evaluated on various architectures, including VITS, language model, and diffusion model-based voice conversion frameworks. These architectures are widely applied in recent speech generation tasks. Especially, compared to the widely used information decoupling methods, our approach offers significant advantages in terms of output quality, computational efficiency, universality, and ease of use, making it promising for extension to other speech generation tasks.

\section{Impact Statement}
This study develops a novel method for extracting timbre-independent content representations, which are crucial for voice conversion tasks that aim to produce speech with specific timbre characteristics. Therefore, it is important to prevent the potential misuse of this technology for fraudulent purposes. For example, through telephonic impersonation, fraudsters can engage in financial swindles, inflicting both pecuniary losses and psychological distress on social members. To counteract the risks of misuse, techniques for detecting converted speech are essential. In the future, we will pay more attention to the research on detecting techniques and the practical applications of such technology.
\bibliography{refs,strings,custom}

\appendix
\onecolumn
\section{Training Details of VITS System}
\label{apx:1}
The VITS systems were trained using the AdamW optimizer \citep{Loshchilov_Hutter_2017} with $\beta_1=0.8$, $\beta_2=0.99$. We set the initial learning rate to $2 \times 10^{-4}$. The learning rate decay was scheduled by a 0.999 factor in every epoch. The models were trained up to 900k steps on 4 NVIDIA V100 GPUs. The batch size was set to 6 per GPU with a maximum segment length of 200 frames.

\section{Training Details of Language Model}
\label{apx:2}
\begin{figure*}[ht]
    \centering
    \includegraphics[scale=0.67]{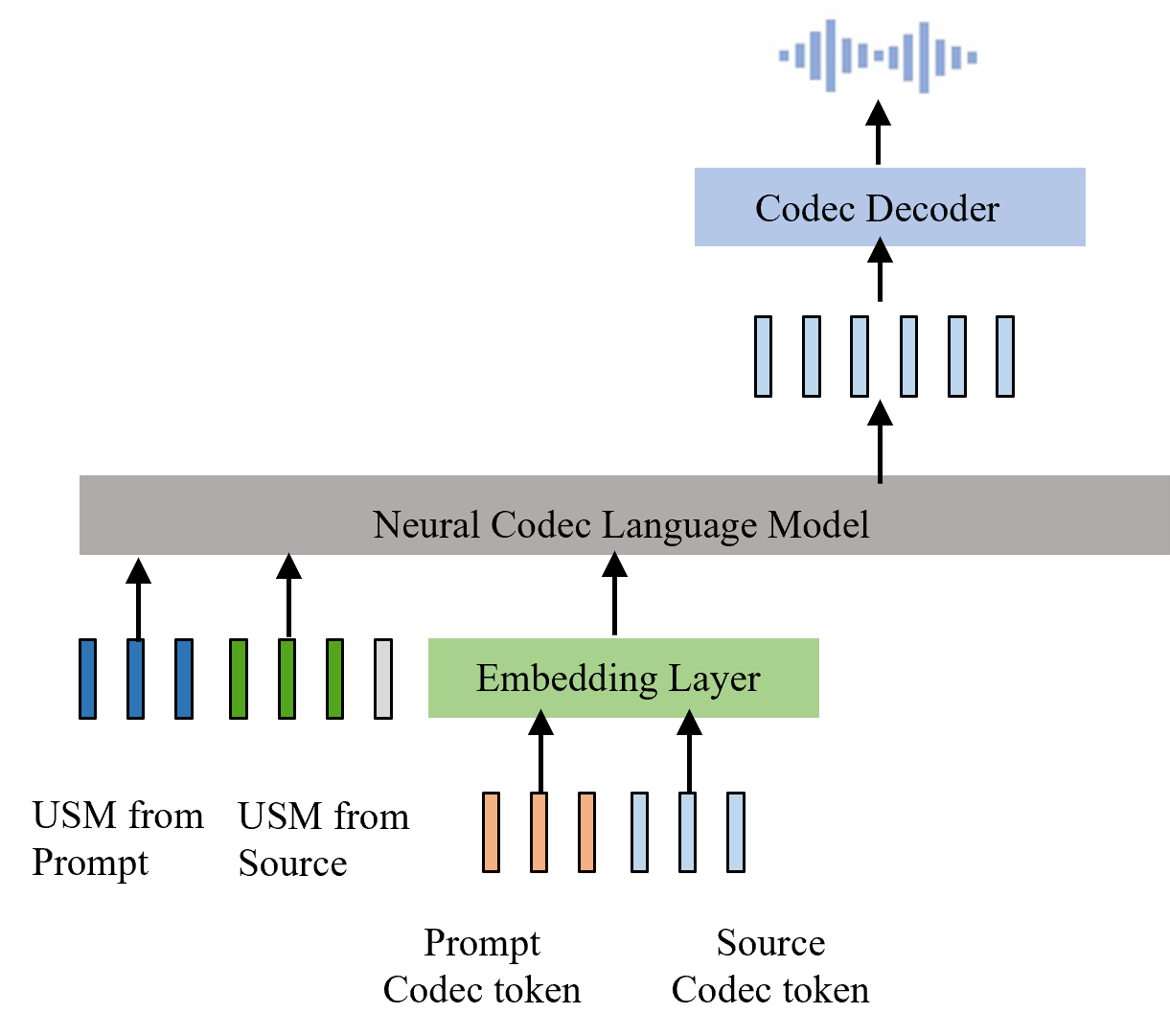}    
    \caption{Illustration of the training for language model based zero-shot voice conversion.}
    \label{LM}
\end{figure*}

Fig.\ref{LM} shows the training process of the language model. Conditioned on the sequence of USM content representations from prompt and source speech, together with the codec tokens from prompt speech segment, the model predicts the target codec tokens hierarchically. The AR and NAR transformers were trained simultaneously using 16 NVIDIA A100 GPUs with a batch size of $2.5$k acoustic tokens per GPU for 150 epochs. We optimize the models with the AdamW optimizer, warm up the learning rate for the first 5k updates to a peak of $2 \times 10^{-4}$.

\section{The Experimental Details of Diffusion Model}
\label{apx:3}

\subsection{The Details of Convolution Transformer in Diffusion Model}

This Convolution Transformer we use consists of 12 Diffusion Convolution Transformer blocks. The detailed configuration of the block is illustrated in Table \ref{DiT_d}. Each block comprises one Film layer, three ConvNeXt blocks, and one DiT block. The Film layer integrates temporal information into the model, while the ConvNeXt blocks and DiT block utilize Adaptive Layer Norm to incorporate speaker embedding information into the model. 

\begin{figure}[htbp]
\centering % 图片居中对齐
\includegraphics[width=0.4\textwidth]{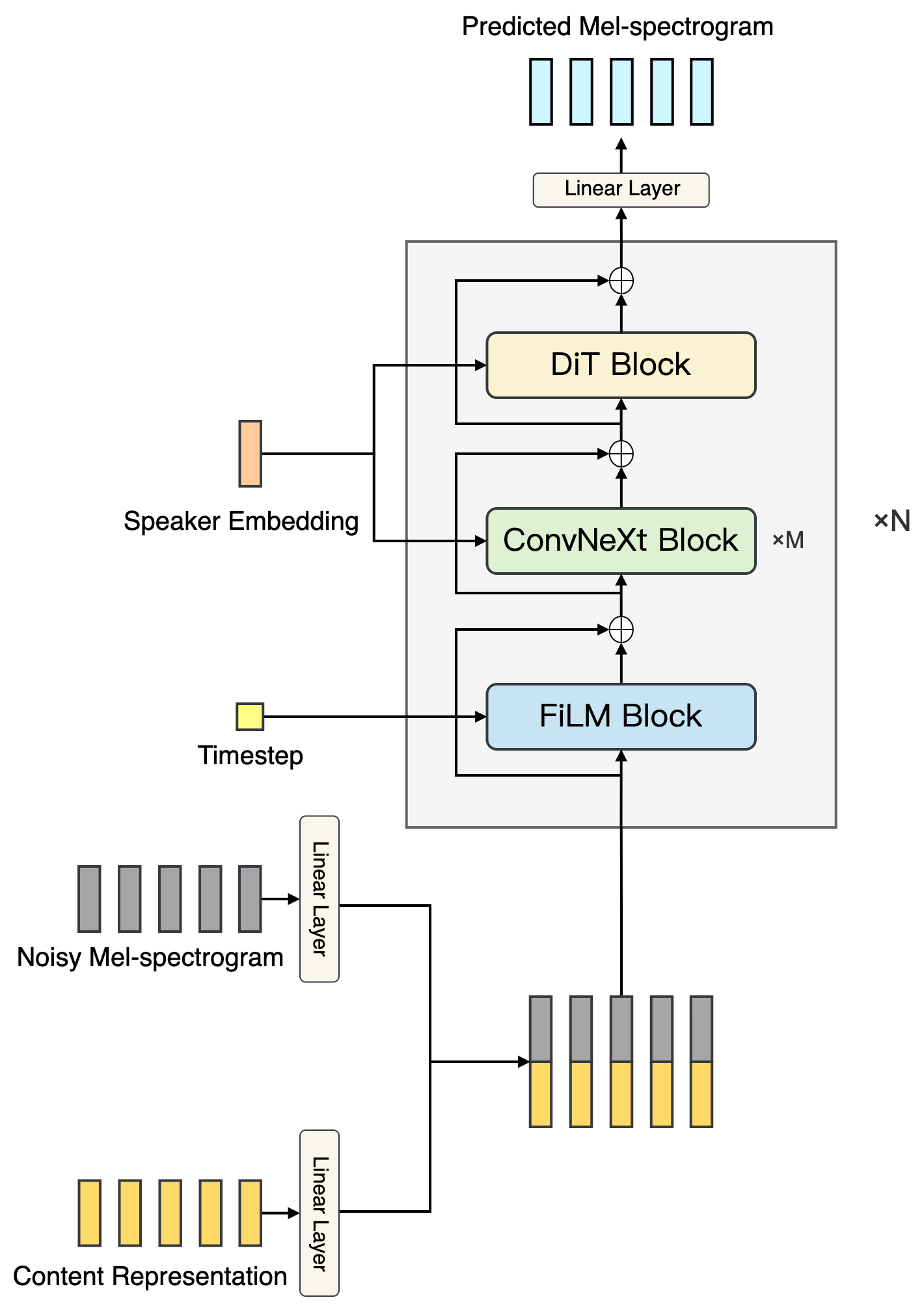}
\caption{Overview of the diffusion model.} 
\label{fig:DiTo} 
\end{figure}

As shown in Fig.\ref{fig:DiTo}. We map both the noisy mel spectrogram and content representation to 384 dimensions and concatenate them together to obtain a 768-dimensional input. We also input both timestep and speaker embedding into the model. Ultimately, we derive an 80-dimensional output aimed at minimizing the loss associated with the diffusion model.

\begin{table}[htbp]
\centering
\begin{tabular}{l l} 
\toprule 
Hyperparameter & Value \\
\midrule
$\rho$ &  7\\
$\sigma_{\min}$ & 0.002 \\
$\sigma_{\max}$ & 80 \\
$\sigma_{data}$ & 0.5 \\
$P_{mean}$ & -1.2 \\
$P_{std}$ & 1.2 \\
$S_{min}$ & 0 \\
$S_{max}$ & infinity \\
$S_{noise}$ & 1 \\
$S_{churn}$ & 0 \\
\bottomrule % 表格底部横线
\end{tabular}
\caption{Diffusion Model Hyperparameters} 
\label{tab:diffusion_hyperparameters} 
\end{table}

\begin{algorithm*}[htbp]
\caption{The sampling process of the diffusion model. Based on Algorithm 2 in \citep{karras2022elucidating}.}
\label{alg:stochastic_sampler}
\begin{algorithmic}[1]
    \REQUIRE $D_{\theta}(x;\sigma,c)$, $t_{i\in\{0,\ldots,N\}}$, $\gamma_{i\in\{0,\ldots,N-1\}}$, $S_{\text{noise}}$, $\textit{cond}$
    \ENSURE $x_{N}$
    \STATE  \textbf{sample} $x_{0}\sim\mathcal{N}(0, t_{0}^{2} I)$
    \FOR{$i\in\{0,\ldots,N-1\}$}
        \STATE  $\gamma_{i}\leftarrow\min\left({S_{\text{churn}}}/{N},\sqrt{2}-1\right)$ if $t_{i}\in[S_{\text{tmin}}, S_{\text{tmax}}]$ else 0
        \STATE  \textbf{sample} $\epsilon_{i}\sim\mathcal{N}(0, S_{\text{noise}}^{2} I)$
        \STATE $\hat{t}_{i}\leftarrow t_{i}+\gamma_{i}t_{i}$         \hfill \COMMENT{Select temporarily increased noise level $\hat{t}_{I}$}
        \STATE  $\hat{x}_{i}\leftarrow x_{i}+\sqrt{\hat{t}_{i}^{2}-t_{i}^{2}}\,\epsilon_{i}$        \hfill \COMMENT{Add new noise to move from $t_{i}$ to $\hat{t}_{i}$}
        \STATE  $d_{i}\leftarrow\left(\hat{x}_{i}-D_{\theta}(\hat{x}_{i};\hat{t}_{i},\textit{cond})\right)/\hat{t}_{i}$     \hfill \COMMENT{Evaluate d$\boldsymbol{x}$/dt at $\hat{t}_{i}$}
        \STATE  $x_{i+1}\leftarrow\hat{x}_{i}+(t_{i+1}-\hat{t}_{i})d_{i}$           \hfill \COMMENT{Take Euler step from $\hat{t}_{i}$ to $t_{i+1}$}
        % \IF{$t_{i+1}\neq 0$}
        %     \STATE  $d_{i}^{\prime}\leftarrow\left(x_{i+1}-D_{\theta}( x_{i+1};t_{i+1})\right)/t_{i+1}$
        %     \STATE  $x_{i+1}\leftarrow x_{i+1}+\left(\frac{1}{2}d_{i}+\frac{1}{2}d_{i}^{\prime}\right)$
        % \ENDIF
    \ENDFOR
    % \RETURN $x_{N}$
\end{algorithmic}
\end{algorithm*}

\subsection{Training and Inference Details of Diffusion Model}
Following CoMoSVC \citep{lu2024comosvc}, we use EDM sampler \citep{karras2022elucidating} as the sampler of diffusion model. We use $D_{\phi}$ to represent the diffusion denoiser. The ground truth mel-spectrogram are denoted as $x_0 \sim p_{data}(x)$ , while the conditional input is denoted by $cond$. The ODE solved by EDM solver can be expressed as follows:
\begin{equation}
   \frac{d x_{t}}{dt} = \frac{x_{t} - D_{\phi}\left(x_{t}, t,\text{cond}\right)}{t},
\end{equation}
where $x_t = x_0 + t \cdot N(0, I)$, represents the ground truth mel-spectrogram corrupted by noise. To make the estimation more flexible, the diffusion decoder $F_{\theta}$, which we use a Diffusion Convolution Transformer, is not applied as the denoiser directly. Instead, A skip connection has been added:
\begin{equation}
\begin{aligned}
% \begin{split}
    D\phi(\mathbf{x}_t, t, \text{cond}) = 
    &c_{\text{skip}}(t)\mathbf{x}_t  
       +c_{\text{out}}(t)\\&F\phi(c_{\text{in}}(t)\mathbf{x}_r, t, c_{\text{noise}}(t)).
% \end{split}
\end{aligned}
\end{equation}

The scaling factors are listed as follows:
\begin{equation}
    c_{\text{skip}}(t) = \frac{\sigma^2_{\text{data}}}{(t - {\sigma_\text{min}})^2 + \sigma^2_{\text{data}}},
\end{equation}
\begin{equation}
    c_{\text{out}}(t) = \frac{\sigma_{\text{data}}(t - {\sigma_\text{min}})}{\sqrt{\sigma^2_{\text{data}} + t^2}},
\end{equation}
\begin{equation}
    c_{\text{int}}(t) = \frac{1}{\sqrt{\sigma^2_{\text{data}} + t^2}},
\end{equation}
\begin{equation}
    c_{\text{noise}}(t) = \frac{1}{4}ln(t).
\end{equation}
The loss function $L_{\phi}$ is:
\begin{equation}
    \mathcal{L}_{\phi}=E\left[\lambda(t)\left\|D_{\phi}\left(x_{t}, t,\text{cond}\right)-x_{0}\right\|^{2}\right],
\end{equation}
where $\lambda(t) = {(t^2 + \sigma^2_{data})}/{(t \cdot \sigma_{data})^2}$, denotes the weight corresponding to different noise levels $t$.

During training, we sample t from $e^{N(P_{min},{P_{{std}}^2})}$. We trained the diffusion model with the AdamW optimizer \citep{Loshchilov_Hutter_2017}, setting $\beta_1=0.9$, $\beta_2=0.999$. We use a initial learning rate of $1 \times 10^{-4}$, which will decay to 90\% of its original value every 100000 steps. We trained the diffusion model on 8 NVIDIA A100 40G GPUs for 185 epochs, each GPU having a batch scale of 40 seconds for 24k waveform.

When performing inference, the timestep sequence $t_0, t_1, ..., t_{n-1}$ is defined as:
\begin{equation}
    t_{i< N} := \left( {\sigma_{\mathrm{max}}}^{\frac{1}{\rho}} + \frac{i}{N-1} \left( {\sigma_{\mathrm{min}}}^{\frac{1}{\rho}} - {\sigma_{\mathrm{max}}}^{\frac{1}{\rho}} \right) \right)^\rho.
\end{equation}
where N is the total sample steps and $\rho$ is the factor that shortens the step lengths near $\sigma_{\min}$ at the expense of longer steps near $\sigma_{\max}$ \citep{li2024styletts}. In order to obtain high quality results, we set $N$ to 30. The rest of the hyperparameter settings are displayed in Table \ref{tab:diffusion_hyperparameters}. Algorithm \ref{alg:stochastic_sampler} demonstrates the sampling process of the diffusion model.

\begin{table*}[htbp]
\centering
\caption{\label{DiT_d}The detailed model configurations of a Convolution Transformer Block.}
\begin{tabular}{@{}lcccc@{}}
\toprule
 Module & Configuration & Value  & Num &\\
\midrule
    \multirow{3}{*}{FiLM Layer}  & Hidden Size & 768 & \multirow{3}{*}{1}\\
    & Conv1D Kernel Size & 1  & \\
    & Conv1D Filter Size & 1536 & \\ 
\midrule
\multirow{5}{*}{ConvNeXt Block}  & Hidden Size & 768 & \multirow{5}{*}{3}\\
    & Conv1D Kernel Size & 7  & \\
    & Conv1D Padding Size & 3  & \\
    & Filter Size & 2048 & \\
\midrule
\multirow{5}{*}{DiT Block}  & Hidden Size & 768 & \multirow{5}{*}{1}\\
    & Attention Heads & 8  & \\
    & Dropout & 0.1 & \\
    & Filter Size & 768 & \\
\bottomrule
\end{tabular}
\end{table*}

\section{Different Weight Combinations on Conversion Effect}
\label{apx:4}
In this section, we investigated the impact of different weight combinations of $w_{1}$ and $w_{2}$ on the conversion performance of different models. As can be observed from the results presented in the Table \ref{weight}, for the VITS-based VC system, when $w_{2}$ is set to 0.2, the system achieves a relatively balanced performance in terms of naturalness and similarity. For larger language model and diffusion model, setting $w_{2}$ at 0.05 yields a balanced performance in both naturalness and similarity. This suggests that larger models are capable of better capturing fine-grained information in representations, such as pronunciation and prosody.
\begin{table*}[htbp]
{
\small
\caption{\label{weight}The impact of different weight combinations on the metrics (UTMOS, SSIM, WER) for different models.}
% \vspace{-s0.5cm}
\begin{center}
\begin{tabular}{c|c|ccc} \hline
System Type & $(w_{1}, w_{2})$ & UTMOS$\uparrow$  & SSIM$\uparrow$ & WER$\downarrow$ \\ \hline
  \multirow{5}{*}{VITS (36M)} & $(1.0,0.0)$  & 3.817 & \textbf{0.806} & 3.918  \\ 
 & $(0.95,0.05)$ & 3.876 & 0.779 & 3.646 \\ 
 & $(0.9,0.1)$ & 3.904 & 0.754 & 2.403 \\ 
 & $(0.8,0.2)$ & 3.937 & 0.748 & 2.102 \\ 
 & $(0.7,0.3)$ & \textbf{3.968} & 0.641 & \textbf{1.997} \\
& $(0.0,1.0)$  & 3.947 & 0.601 &  2.285 \\ \hline
\multirow{4}{*}{Language Model (227M)} & $(1.0,0.0)$  & 3.906 & \textbf{0.772}  & 2.315 \\ 
 & $(0.95,0.05)$  & 4.011 & 0.751 & 2.133 \\ 
 & $(0.9,0.1)$  & \textbf{4.065} & 0.723 & \textbf{1.989} \\
 & $(0.8,0.2)$  & 4.054 & 0.679 &  2.012 \\ 
 & $(0.0,1.0)$  & 4.058 & 0.641 &  2.153 \\ \hline
 \multirow{2}{*}{Diffusion Model (287M)} & $(1.0,0.0)$  & 3.678 & \textbf{0.793}  & 3.185 \\
 & $(0.95,0.05)$  & \textbf{3.791} & 0.759 &  1.575 \\ 
 & $(0.0,1.0)$  & 3.773 & 0.652 &  \textbf{1.338} \\ \hline
\end{tabular}

\end{center}}
%\vspace{-0.5cm}
\end{table*}

\end{document}